\documentclass[
    ,final            % use final for the camera ready runs
  ]
  {aipproc}

\layoutstyle{6x9}

\begin{document}

\title{Stellar spindown: From the ONC to the Sun}


\author{Alexander Scholz}{
  address={SUPA, School of Physics \& Astronomy, University of St. Andrews, North Haugh, St. Andrews,
  Fife KY16 9SS, United Kingdom, email: as110@st-andrews.ac.uk}
}

\begin{abstract}
Rotation is a key parameter in the evolution of stars. From 1 Myr (the age of the ONC) to 4.5 Gyr 
(the age of the Sun), solar-like stars lose about 1-2 orders of specific angular momentum. The main 
agents for this rotational braking are believed to be star-disk interaction and magnetically powered 
stellar winds. Over the last decade, the observational fundament to probe the stellar spindown has 
dramatically improved. Significant progress has been made in exploring the underlying physical 
causes of the rotational braking. Parameterized models combining the effects of star-disk interaction, 
winds, and pre-main sequence contraction are able to reproduce the main features of the rotational 
data for stars spanning more than 3 orders of magnitude in age. This has allowed us to constrain 
stellar ages based on the rotation rates ('gyrochronology'). One main challenge for future work 
is to extend this type of analysis to the substellar mass range, where the rotational database is 
still sparse. More theoretical and observational work is required to explore the physics of the braking 
processes, aiming to explain rotational evolution from first principles. In this review for Cool Stars
15, I will summarize the status quo and the recent developments in the field.
\end{abstract}

\maketitle

%%%%%%%%%%%%%%%%%%%%%%%%%%%%%%%%%%%%%%%%%%%%
%% MAINMATTER
%%%%%%%%%%%%%%%%%%%%%%%%%%%%%%%%%%%%%%%%%%%%

\section{Introduction}

Analysing the spindown of low-mass stars is one of the most notorious and difficult issues in astrophysics. 
We are dealing with a multi-faceted problem: The spin of stars is a function to fundamental stellar properties 
-- mass, radius, age. The rotation rate is also affected by the star formation process, including 
the collapse of molecular cloud cores, the formation of binaries and planetary systems, and accretion 
from circumstellar disks. Finally, the stellar magnetic field, its generation, structure and interaction 
with the stellar atmosphere plays a dominant role for the regulation of the rotation from the T Tauri 
phase to the age of the Sun. A comprehensive discussion of the spindown of stars thus requires input 
from a broad range of fields. While I will mostly focus on rotation itself, I encourage the reader 
to explore related chapters in this volume, for example the reviews by Edwards, Donati, K{\"u}ker, 
Stelzer, Basri, as well as the splinter session summary by Reiners et al..

In this review, I will trace the spin of stars from $\sim 1$\,Myr, the age of the Orion Nebula Cluster,
to 4.5\,Gyr, the age of the Sun. In this time frame, the angular momentum of stars is primarily 
controlled by two mechanisms: 1) {\it Disk braking}, used here as a generic term for
mechanisms that remove angular momentum from the star via the interaction between star and disk, 
including disk-locking and accretion-powered stellar winds. 2) {\it Wind braking}, i. e. angular 
momentum losses due to magnetized stellar winds.

While disk braking operates on relatively short timescales compatible with the disk lifetime 
of 1-10\,Myr, wind braking is a long-term process and regulates the spindown on timescales 
of hundreds of Myr. This allows us to discuss the two processes separately: In the first part 
of this review, the focus will be on disk braking, i.e. the evolution in the pre-main sequence phase. 
In the second part, wind braking will be discussed by investigating the 
long-term evolution on the main-sequence. I will focus on objects with masses 
ranging from about one solar mass down to the substellar regime.

Reviewing the rotational evolution of stars and brown dwarfs is a timely exercise, for two reasons:

1) We have recently experienced an enormous growth in the number of objects with known
periods, covering a wide range of ages and masses. Rotation rates are traditionally measured using
two techniques, photometric monitoring providing rotation periods and high-resolution spectroscopy providing
projected rotational velocities $v\sin{i}$ from Doppler line broadening. Thanks to the availability 
of wide-field imagers at 2-4\,m class telescopes as well as the exploitation of planetary 
transit survey, the rotational database now comprises periods for about 3000 objects, and counting.
In addition, the availability of high-resolution multi-object spectrographs (e.g., FLAMES at the
VLT) has significantly enlarged the pool of $v\sin{i}$ data.

2) A number of new developments over the past five years have enabled studies that
provide important complementary insights. Examples are the availability of the Spitzer
telescope for mid-infrared studies of disk braking (see contribution by Baliber, this volume)
and the ongoing efforts to determine magnetic field structures from Zeeman Doppler Imaging 
(see review by Donati, this volume). Combining the rotational data with these additional
information yields a new, comprehensive picture of the angular momentum evolution of stars.

\begin{figure}
  \includegraphics[height=.5\textheight,angle=-90]{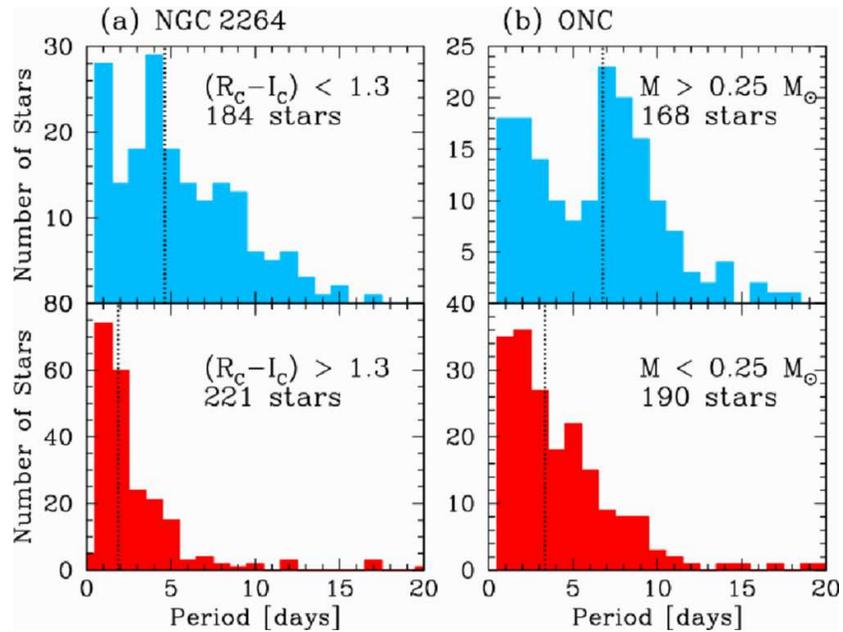}
  \caption{Distribution of rotation periods for stars in NGC2264 (age $\sim 2$\,Myr, left panels) 
  and the ONC (age $\sim 1$\,Myr, right panels). The samples are split in two mass bins at roughly
  $M\sim 0.25\,M_{\odot}$. High and low mass stars are shown in the upper and lower panels, 
  respectively. Figure from \citet{2005A&A...430.1005L}.
  \label{f1}}
\end{figure}

\section{Pre-main sequence evolution}
\label{pms}

\subsection{Initial period distribution}

The initial distribution of rotation rates of low-mass stars is now well-established, thanks
to extensive survey work in clusters at 1-2\,Myr, mainly the ONC and NGC2264. For a detailed 
discussion of the available data in these two clusters see for example 
\citet{1999AJ....117.2941S,2001ApJ...554L.197H,2002A&A...396..513H,2001AJ....121.1676R,2004AJ....127.2228M,2004A&A...417..557L,2005A&A...430.1005L}.
Fig. \ref{f1} shows the period distribution in the ONC and NGC2264, as published by 
\citet{2005A&A...430.1005L}. Three features in these histograms should be highlighted: a) At 
this early age, the objects show a broad range of periods, mostly between 1 and 10\,d with a 
tail extending to $\sim 20$\,d. b) The median period drops with decreasing object mass in the 
considered mass range. c) The distribution is bimodal for solar-mass stars, but unimodal for 
very low mass stars (see the review by \citet{2007prpl.conf..297H}).

\begin{figure}
  \includegraphics[height=.5\textheight,angle=-90]{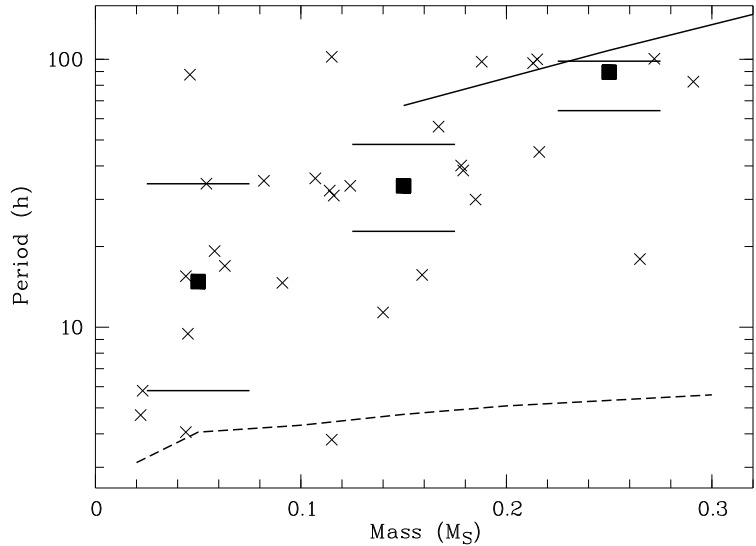}
  \caption{Rotation periods for very low mass stars and brown dwarfs in the young population around
  $\epsilon$\,Ori. Large symbols denote the median period, horizontal solid lines the quartiles per 0.1$\,M_{\odot}$ 
  mass bin. The solid line in the upper right corner shows the period-mass relation in the ONC, as given by
  \citet{2001ApJ...554L.197H}. The dashed line is the approximate breakup limit for an age of 5\,Myr. 
  Figure from \citet{2005A&A...429.1007S} \label{f2}}
\end{figure}

As of today, the datasets in the ONC and NGC2264 do not include the lowest mass stars and brown
dwarfs (but see the contribution by Rodriguez-Ledesma in this volume for an update). In general,
the period database for substellar objects is sparsely populated. Nonetheless, the available
data allows us first important conclusions. In Fig. \ref{f2} we show periods for objects with
masses of 0.02 to 0.3$\,M_{\odot}$ in the $\sim 5$\,Myr old $\epsilon$\,Ori region. 
As can be seen in this plot, the trend of declining median period towards lower masses, as seen
in the ONC, continues steadily in the substellar regime. As a result, the median period for
brown dwarfs is shorter than one day, and the fast rotators in this mass regime are at
periods of 3-5\,h (\citet{2001A&A...367..218B,2003A&A...408..663Z,2004A&A...419..249S,2005A&A...429.1007S}).
These rotation rates are comparable to the {\it breakup limit}, the value where centrifugal and 
gravitational forces are in balance at the equator.

\subsection{Disk braking at 1-5\,Myr}

\begin{figure}
  \includegraphics[height=.44\textheight]{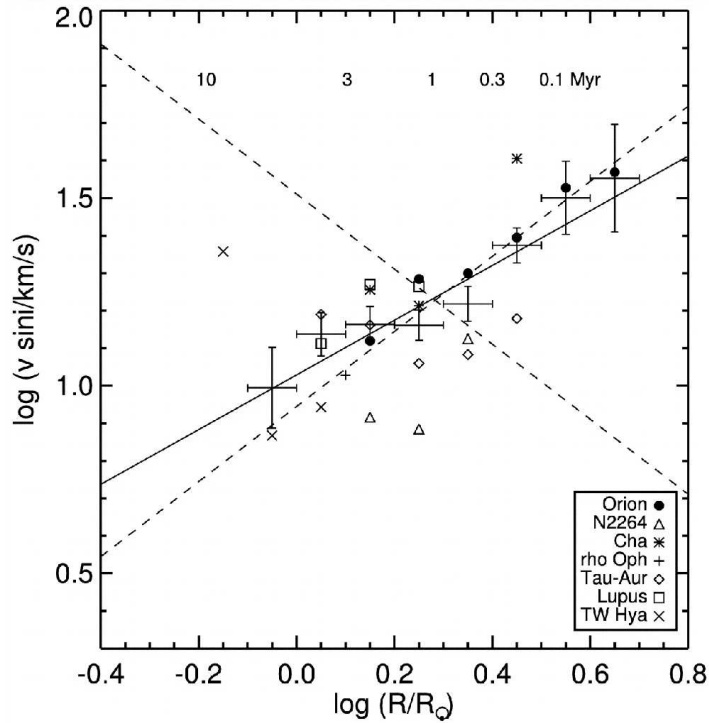}
  \caption{Average $\log{(v \sin{i})}$  vs. $\log{R}$ for young stars. The points with error bars 
  in both directions are the average $\log{(v \sin{i})}$ for all stars within the specified range 
  in $\log{R}$. The $\log{(v \sin{i})}$ values for many individual clusters are also plotted,
  Approximate ages as a function of R are indicated. The solid line shows the best linear fit,
  exluding the datapoint at 10\,Myr. The dashed lines are the relations expected for evolution 
  with constant angular velocity (slope of $1$), and constant angular momentum (slope of $-1$). 
  The observed slope is within 2$\sigma$ of the prediction for constant angular velocity, but
  inconsistent with the value expected for conservation of angular momentum. 
  Figure from \citet{2004AJ....127.1029R} \label{f3}}
\end{figure}

The first 5\,Myr in the spin evolution are characterized by strong rotational regulation. This has 
been firmly established in a number of recent studies, based on the growing rotational database in 
young open clusters and star forming regions (e.g.
\citet{2002ApJ...564..877T,2002AJ....124..546R,2004AJ....127.1029R,2005A&A...430.1005L,2005ApJ...633..967H},
see also the review in \citet{2007prpl.conf..297H}). 
A good illustration of this finding is given in Fig. \ref{f3}: The average $v\sin{i}$ at 1-5\,Myr is 
clearly inconsistent with angular momentum conservation;
instead it roughly follows the trend expected for angular velocity (hence period) conservation. 
This can be understood as a consequence of a rotational braking mechanism affecting a fraction
of the objects that decreases with age. The maximum timescale for this type of regulation is 
$\sim 5$\,Myr, consistent with the lifetime of accretion disks, a first indication that 
the braking mechanism is related to the presence of a disk. 

In fact, this is the underlying assumption in the most commonly discussed scenarios for the strong
rotational regulation in the T Tauri phase (called 'disk braking' in the following). The theories
for this mechanism essentially fall in two groups: Either the angular momentum is extracted directly
from the star, or it is transferred from the star to the disk. 'Disk-locking' is probably the leading idea 
here: According to the standard picture (\citet{1990RvMA....3..234C,1991ApJ...370L..39K}), star and disk are 
coupled by the strong stellar magnetic field. As a result, the disk exhibits a torque onto the star, and thus 
slows down its rotation. Angular momentum is moved from the star to the disk and then carried away in a disk 
wind. One possible realization of a disk-locking scenario is the X-wind model 
(e.g., \citet{1994ApJ...429..781S,2008arXiv0806.4769M}). 

Alternative models for disk braking are based on stellar winds. A solar-type wind as in main-sequence
objects, however, would not be sufficient for the strong braking seen in T Tauri stars. This leads to
the idea of stellar winds powered by accretion. \citet{2005ApJ...632L.135M} have demonstrated that accretion-driven
winds can in principle explain the observational picture, if the mass outflow rates are a substantial fraction 
of the accretion rates ($\sim 10$\%).

For observational studies of disk braking, it is important to point out that both types of models -- 
accretion-powered winds and disk-locking -- require the presence of a disk and a coupling between
star and disk. Thus, observations should primarily aim to establish whether a connection between
disk and rotation is present or not. \citet{1993AJ....106..372E} have presented first evidence for such a 
connection, in the sense that objects with near-infrared colour-excess, indicative of an inner disk, 
are primarily seen as slow rotators. While some studies have confirmed this early finding (e.g., 
\citet{2002A&A...396..513H}), others did not (e.g., \citet{2001AJ....121.1003S}). 

The mid-infrared observations from the Spitzer Spacecraft Telescope, launched 2003, have greatly improved our 
understanding of circumstellar disks. For the first time, Spitzer provides unambiguous and reliable 
disk indicators for the large numbers of objects in the ONC and NGC2264 with measured rotation periods. 
Based on Spitzer data, \citet{2006ApJ...646..297R} indeed find strong evidence for disk braking:
The overwhelming majority of the objects with disks are slow rotators with periods $>1.8$\,d. 
A similarly clear result has been obtained for NGC2264 by \citet{2007ApJ...671..605C}. It is probably
fair to say that most the researchers working in the field today believe that the presence of a disk 
clearly plays a role for the rotational braking.

On the other hand, negative results regarding disk braking have been obtained for IC348 (age 2-4\,Myr, 
\citet{2006ApJ...649..862C}), Taurus, and Chamaeleon I ($\sim 2$\,Myr, Nguyen et al., in prep.). 
We are still looking at partly controversial findings, possibly pointing at environmental differences. 
A number of aspects can dilute a signature of disk braking, e.g., age spread, binarity, dispersion in
disk lifetimes, or insufficient sample size. If and how these factors can explain all the observational results 
is currently under investigation (see the contribution by Baliber, this volume). In standard disk braking 
scenarios it is not the mere presence of a disk, but the coupling between star and disk that provides the 
angular momentum removal. Thus, a more detailed understanding requires to look at a variety of disk 
braking diagnostics, including emission lines affected by accretion and winds (see review by Edwards, 
this volume).
 
Another open issue is the mass dependence of this mechanism, which can be probed by looking at the
very low mass objects in star forming region. Studies of disk braking in
brown dwarfs are rare, mainly due to a lack of overlap of period and Spitzer data. 
The few existing constraints, however, indicate that disk braking is at work
in substellar objects as well (\citet{2004A&A...419..249S,2005MmSAI..76..303M}). However, the signature 
of disk braking, as seen in Spitzer data, becomes weaker at very low masses (\citet{2006ApJ...646..297R}), 
pointing to 'imperfect' disk braking (\citet{2005A&A...430.1005L}). This might be explained by a mass 
dependency in the magnetic field topology and will be subject of future programs.

\section{Main-sequence evolution}
\label{ms}

\subsection{The pre-main sequence transition}

The rotational picture in the age range 10-100\,Myr is difficult to interpret, because we
see superimposed the effects of disk braking, the strong pre-main sequence contraction, as well
as the beginning of wind braking. In a recent paper, however, we have shown that the median
$v\sin{i}$ for stars covering ages from 5 to 30\,Myr is inconsistent with the expected
evolution for angular velocity conservation, but roughly in agreement with angular momentum conservation 
(\citet{2007ApJ...662.1254S}). This is in stark contrast to the behaviour at 1-5\,Myr, 
as discussed above (see Fig. \ref{f3}), indicating a sharp change in the rotational braking
at 5-10\,Myr. This is readily explained by the disk dissipation, causing the breakdown of
disk braking, resulting in rapid spinup due to the contraction.

Including rotational data at $\sim 100$\,Myr shows the onset of wind braking: The median rotation 
rates are significantly lower than expected from pure angular momentum conservation (\citet{2007ApJ...662.1254S}). 
Wind braking is thought to be the dominant mechanism of rotational braking on the main-sequence and 
will be discussed in detail below. This sequence of events -- disk braking, rapid spinup due to 
contraction, and onset of wind braking -- is probably applicable to very low mass stars 
and brown dwarfs as well (\citet{2004A&A...421..259S}, Scholz \& Eisl{\"o}ffel, in prep.).

\subsection{Wind braking: theory vs. observations}

The rotational braking due to stellar winds is usually quantified in parameterized 
angular momentum loss laws based on the description given by \citet{1988ApJ...333..236K},
derived from wind physics by \citet{1984LNP...193...49M}:
$\frac{dJ}{dt} \propto \omega ^{x} R^{0.5} M^{-0.5}$. We distinguish two 
regimes: 

a) The linear regime includes all objects with $\omega < \omega_{\mathrm{crit}}$ 
and is calculated with $x=3$, reproducing the Skumanich law $J \propto t^{-0.5}$,
the empirically established angular momentum loss law for solar-type main-sequence
stars (\citet{1972ApJ...171..565S}). 

b) The saturated regime comprises all fast rotating 
objects with $\omega > \omega_{\mathrm{crit}}$ and is characterized by $x=1$, resulting 
in an exponential evolution of angular momentum. To fit the observational data, 
$\omega_{\mathrm{crit}}$ is normally assumed to be a function of object mass. This
semi-empirical description is at the core of the most commonly used models for 
the rotational evolution 
(\citet{1997ApJ...480..303K,1997A&A...326.1023B,2000ApJ...534..335S,2003ApJ...586..464B,2008ApJ...684.1390R}).
For the two different regimes, I will use the nomenclature suggested by \citet{2003ApJ...586..464B}: 
I-sequence for the linear regime and C-sequence for the saturated regime.

\begin{figure}
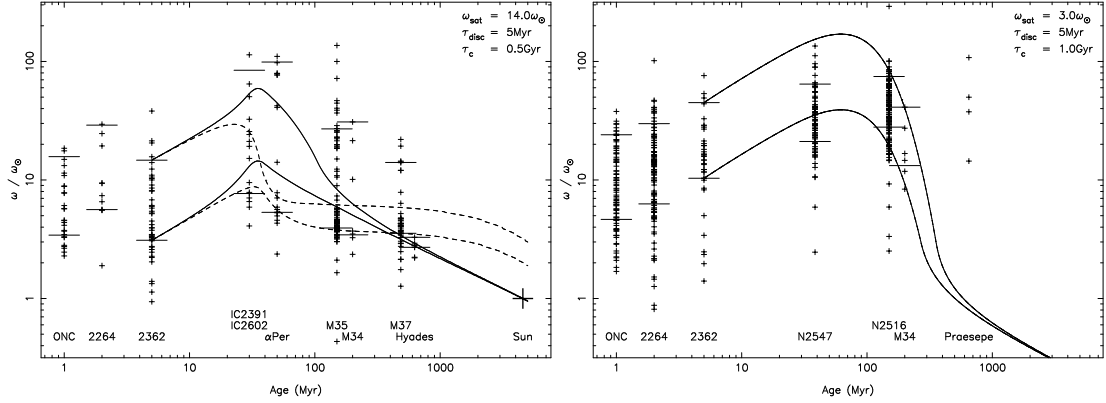

  \includegraphics[height=.33\textheight,angle=-90]{f9a.ps}
  \includegraphics[height=.33\textheight,angle=-90]{f9b.ps}
  \caption{Rotation periods from the Monitor project and other literature sources for
  0.9-1.1$\,M_{\odot}$ (left panel) and 0.2-0.35$\,M_{\odot}$ (right panel). The horizontal solid lines
  show the typical range of the periods in each cluster (measured as 25\% and 90\% percentiles).
  The evolutionary tracks shown in solid lines are for solid-body rotation, dashed line include
  core-envelope decoupling.
  The figures are updated versions of the ones shown in \citet{2008MNRAS.383.1588I}, including new 
  data in M35, M37, and Praesepe (\citet{2008arXiv0803.1488H,2008arXiv0805.1040M,2007MNRAS.381.1638S},  
  figure kindly provided by J. Irwin).\label{f9}}
\end{figure}

In Fig. \ref{f9} a large sample of rotation periods, mostly from the {\it Monitor} program
(\citet{2006MNRAS.370..954I,2007MNRAS.377..741I,2008MNRAS.383.1588I,2008MNRAS.384..675I}), is
compared to models including disk braking at 1-5\,Myr and wind 
braking following the approach given above. The models provide a reasonable fit to the 
typical period evolution for solar-mass stars from a few Myr to 4.5\,Gyr. For very low masses,
the models reproduce the datapoints for ages $<500$\,Myr, but require a different scaling for 
$\omega_{\mathrm{crit}}$ to match the periods in Praesepe. Most solar-mass stars 
go through a phase on the C-sequence at ages $<100$\,Myr, before they switch to the I-sequence, 
causing rapid spindown. In contrast, very low mass stars stay on the C-sequence for timescales 
$>100$\,Myr and thus continue to be fast rotators. As is evident from Fig. \ref{f9},
there are clearly issues in reproducing the total range of periods at any given age, which 
can partly be resolved by taking into account core-envelope decoupling, variable disk lifetimes, 
and binarity. 

The availability of parameterized models for the rotational evolution allows us to 
pursue 'gyrochronology' -- measuring ages of stars based on rotation rates -- a method recently
put forward by \citet{2007ApJ...669.1167B} and others. Gyrochronology relies on an accurate
calibration of the spindown rates, obtained from stars with known ages. While gyro ages for solar-mass 
stars may already be more reliable than ages estimated from other indicators (e.g., activity, kinematic),
the calibration at very low masses requires further work.

\subsection{The bimodal spindown} 

\begin{figure}
  \includegraphics[height=.5\textheight,angle=-90]{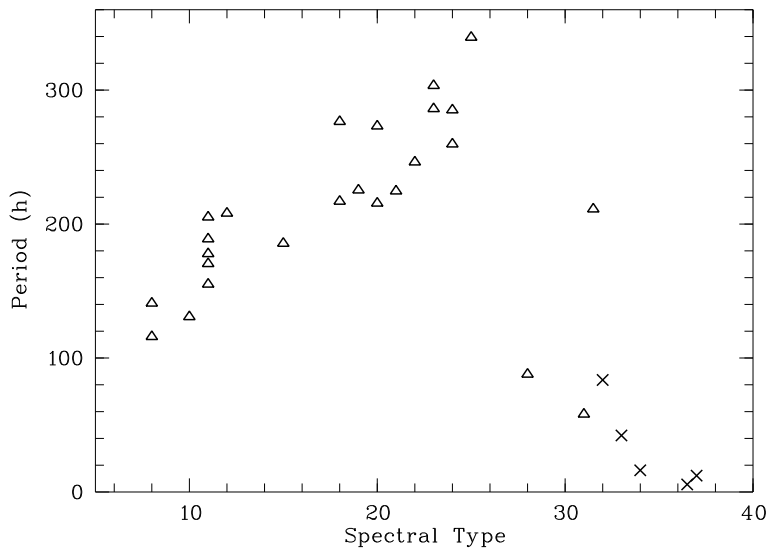}
  \caption{Rotation periods in clusters at ages $\sim 700$\,Myr as a function of spectral type. The
  plot includes periods in the Hyades from \citet{1987ApJ...321..459R,1995PASP..107..211P}, shown as 
  triangles, and periods in Praesepe from \citet{2007MNRAS.381.1638S}, shown as crosses. The spectral 
  type is parameterized as follows: 10 -- G0, 20 -- K0, 30 -- M0. \label{f7}}
\end{figure}

As discussed above, the common description of the main-sequence wind braking includes a
bimodal angular momentum loss law. This bimodality can be clearly seen
in observational data, particularly when rotation is plotted vs. object mass. In Fig. \ref{f7}
we show periods in clusters at an age of $\sim 700$\,Myr as a function of spectral type,
used as a proxy for mass (\citet{2007MNRAS.381.1638S}). At this age, the objects have essentially
'forgotten' their pre-main sequence history, thus their rotation rates are primarily determined by
wind braking.

The most obvious feature in the period-mass relation is a break at spectral types K8-M2, 
corresponding to masses of $\sim 0.5\,M_{\odot}$. While late K stars are already spun down at 
the age of 700\,Myr to periods $>10$\,d, M stars at the same age are still fast rotating with 
periods $<4$\,d. The figure demonstrates that the decline of the periods continues 
to late M spectral types. The transition from slowly rotating K dwarfs (I-sequence) to fast 
rotating M dwarfs (C-sequence) separates the two regimes of rotational braking. This transition 
is likely to be a function of mass, shifting to later spectral types as the objects become older. 
In $v\sin{i}$ data of field dwarfs it is observed at spectral types of $\sim$M3-M4 
(\citet{1998A&A...331..581D}, see Reiners et al., this volume).

Interpreting the transition from I- to C-sequence and the underlying physics is one of the most 
relevant outstanding issues in this field. As of today, we do not have a clear understanding for the 
breakdown of the Skumanich law at very low masses. It is tempting to explain
the bimodal spindown as a consequence of the interior structure of stars: While solar-mass stars
on the main-sequence have a radiative core, pre-main sequence stars and very low mass objects are fully
convective. The presence of an interface layer between radiative core and convective envelope is
often argued to be of prime importance for the generation of the large-scale magnetic field of
the Sun (see also the contributions by Priest, K{\"u}ker, this volume). The absence of a radiative
core may cause a change in the magnetic field generation, the magnetic topology, and thus the
characteristics of the stellar wind.

There is indeed evidence for a change in the magnetic properties that goes along with the breakdown
of the Skumanich braking at very low masses. For example, H$\alpha$ activity is strongly enhanced
at spectral types $>$M3 in field objects (\citet{1998A&A...331..581D}). Moreover, there are indications 
that the structure of the large-scale magnetic field (\citet{2006Sci...311..633D}) as well as the properties 
of magnetically induced spots (\citet{2005A&A...438..675S}) change in the very low mass regime. Thus, 
the observed change in the spindown at very low masses may be one observational manifestation among 
others for the presence of two regimes of magnetic properties. 

For further progress in this field, it is crucial to learn more about dynamo activity, magnetic
field emergence, stellar winds, and how they depend on stellar age and mass. In particular, the currently 
used parameterized wind braking law as described above is suspect, as it has been designed and repeatedly 
adapted to account for observational findings, for example to match the Skumanich law and to reproduce 
fast rotating very low mass objects (e.g., \citet{2000ApJ...534..335S}). Studies on how to improve the 
currently existing models are underway (e.g., \cite{2008ApJ...678.1109M}). The long-term perspective
should be to replace the parameterized spindown laws by physical models, and thus describe the
rotational evolution from first principles.

%%%%%%%%%%%%%%%%%%%%%%%%%%%%%%%%%%%%%%%%%%%%%%%%
%% BACKMATTER
%%%%%%%%%%%%%%%%%%%%%%%%%%%%%%%%%%%%%%%%%%%%%%%%

\begin{theacknowledgments}
  I thank the SOC and LOC of Cool Stars 15 for organising an exciting and
  instructive conference. For helpful discussions regarding issues discussed 
  in this paper, I am grateful to Ansgar Reiners, Jerome Bouvier, Sean Matt, 
  Jonathan Irwin, Andrew Collier Cameron, Nairn Baliber, and Duy C. Nguyen. I would like to 
  thank Jochen Eisl{\"o}ffel for nearly a decade of fruitful collaboration on 
  stellar rotation.
\end{theacknowledgments}



\bibliographystyle{aipproc}   % if natbib is available

%\newcommand\aj{\ref@jnl{AJ}} %% Astronomical Journal
%\newcommand\apj{\ref@jnl{ApJ}} %% Astrophysical Journal

%\bibliography{aleksbib}

\begin{thebibliography}{9}

\bibitem[Bailer-Jones 
\& Mundt(2001)]{2001A&A...367..218B} Bailer-Jones, C.~A.~L., \& Mundt, R.\ 2001, Astronomy \& Astrophysics, 
367, 218 

\bibitem[Barnes(2007)]{2007ApJ...669.1167B} Barnes, S.~A.\ 2007, Astrophysical Journal, 669, 
1167 

\bibitem[Barnes(2003)]{2003ApJ...586..464B} Barnes, S.~A.\ 2003, Astrophysical Journal, 586, 
464 

\bibitem[Bouvier et 
al.(1997)]{1997A&A...326.1023B} Bouvier, J., Forestini, M., \& Allain, S.\ 1997, Astronomy \&
Astrophysics, 326, 1023 

\bibitem[Camenzind(1990)]{1990RvMA....3..234C} Camenzind, M.\ 1990, Reviews 
in Modern Astronomy, 3, 234 

\bibitem[Cieza 
\& Baliber(2007)]{2007ApJ...671..605C} Cieza, L., \& Baliber, N.\ 2007, Astrophysical 
Journal, 671, 605 

\bibitem[Cieza 
\& Baliber(2006)]{2006ApJ...649..862C} Cieza, L., \& Baliber, N.\ 2006, Astrophysical
Journal, 649, 862 

\bibitem[Delfosse et 
al.(1998)]{1998A&A...331..581D} Delfosse, X., Forveille, T., Perrier, C., \& Mayor, M.\ 1998, Astronomy
\& Astrophysics, 331, 581 

\bibitem[Donati et al.(2006)]{2006Sci...311..633D} Donati, J.-F., 
Forveille, T., Cameron, A.~C., Barnes, J.~R., Delfosse, X., Jardine, M.~M., 
\& Valenti, J.~A.\ 2006, Science, 311, 633 

\bibitem[Edwards et al.(1993)]{1993AJ....106..372E} Edwards, S., et al.\ 
1993, Astronomical Journal, 106, 372 

\bibitem[Hartman et al.(2008)]{2008arXiv0803.1488H} Hartman, J.~D., et al.\ 
2008, Astrophysical Journal, submitted, arXiv:0803.1488 

\bibitem[Herbst et al.(2007)]{2007prpl.conf..297H} Herbst, W., 
Eisl{\"o}ffel, J., Mundt, R., 
\& Scholz, A.\ 2007, Protostars and Planets V, 297 

\bibitem[Herbst 
\& Mundt(2005)]{2005ApJ...633..967H} Herbst, W., \& Mundt, R.\ 2005, Astrophysical Journal, 
633, 967 

\bibitem[Herbst et 
al.(2002)]{2002A&A...396..513H} Herbst, W., Bailer-Jones, C.~A.~L., Mundt, R., Meisenheimer, K., 
\& Wackermann, R.\ 2002, Astronomy \& Astrophysics, 396, 513 

\bibitem[Herbst et al.(2001)]{2001ApJ...554L.197H} Herbst, W., 
Bailer-Jones, C.~A.~L., \& Mundt, R.\ 2001, Astrophysical Journal Letters, 554, L197 

\bibitem[Irwin et al.(2008)]{2008MNRAS.384..675I} Irwin, J., Hodgkin, S., 
Aigrain, S., Bouvier, J., Hebb, L., Irwin, M., 
\& Moraux, E.\ 2008, Monthly Notices of the Royal
Astronomical Society, 384, 675 

\bibitem[Irwin et al.(2008)]{2008MNRAS.383.1588I} Irwin, J., Hodgkin, S., 
Aigrain, S., Bouvier, J., Hebb, L., \& Moraux, E.\ 2008, Monthly Notices of the Royal
Astronomical Society, 383, 1588 

\bibitem[Irwin et al.(2007)]{2007MNRAS.377..741I} Irwin, J., Hodgkin, S., 
Aigrain, S., Hebb, L., Bouvier, J., Clarke, C., Moraux, E., 
\& Bramich, D.~M.\ 2007, Monthly Notices of the Royal
Astronomical Society, 377, 741 

\bibitem[Irwin et al.(2006)]{2006MNRAS.370..954I} Irwin, J., Aigrain, S., 
Hodgkin, S., Irwin, M., Bouvier, J., Clarke, C., Hebb, L., 
\& Moraux, E.\ 2006, Monthly Notices of the Royal
Astronomical Society, 370, 954 

\bibitem[Kawaler(1988)]{1988ApJ...333..236K} Kawaler, S.~D.\ 1988, Astrophysical Journal, 
333, 236 

\bibitem[Koenigl(1991)]{1991ApJ...370L..39K} Koenigl, A.\ 1991, Astrophysical Journal Letters, 370, 
L39 

\bibitem[Krishnamurthi et al.(1997)]{1997ApJ...480..303K} Krishnamurthi, 
A., Pinsonneault, M.~H., Barnes, S., \& Sofia, S.\ 1997, Astrophysical Journal, 480, 303 

\bibitem[Lamm et al.(2005)]{2005A&A...430.1005L} Lamm, M.~H., Mundt, R., Bailer-Jones, C.~A.~L., 
\& Herbst, W.\ 2005, Astronomy \& Astrophysics, 430, 1005 

\bibitem[Lamm et 
al.(2004)]{2004A&A...417..557L} Lamm, M.~H., Bailer-Jones, C.~A.~L., Mundt, R., Herbst, W., 
\& Scholz, A.\ 2004, Astronomy \& Astrophysics, 417, 557 

\bibitem[Makidon et al.(2004)]{2004AJ....127.2228M} Makidon, R.~B., Rebull, 
L.~M., Strom, S.~E., Adams, M.~T., \& Patten, B.~M.\ 2004, Astronomical Journal, 127, 2228 

\bibitem[Matt 
\& Pudritz(2008)]{2008ApJ...678.1109M} Matt, S., \& Pudritz, R.~E.\ 2008, Astrophysical Journal, 678, 1109 

\bibitem[Matt 
\& Pudritz(2008)]{2008ASPC..384..339M} Matt, S., \& Pudritz, R.\ 2008, 14th Cambridge Workshop on 
Cool Stars, Stellar Systems, and the Sun, 384, 339 

\bibitem[Matt 
\& Pudritz(2005)]{2005ApJ...632L.135M} Matt, S., \& Pudritz, R.~E.\ 2005, Astrophysical Journal Letters, 632, 
L135 

\bibitem[Meibom et al.(2008)]{2008arXiv0805.1040M} Meibom, S., Mathieu, 
R.~D., \& Stassun, K.~G.\ 2008, Astrophysical Journal, submitted, arXiv:0805.1040 

\bibitem[Mestel(1984)]{1984LNP...193...49M} Mestel, L.\ 1984, Cool Stars, 
Stellar Systems, and the Sun, 193, 49 

\bibitem[Mohanty 
\& Shu(2008)]{2008arXiv0806.4769M} Mohanty, S., \& Shu, F.~H.\ 2008, Astrophysical Journal, in press,
arXiv:0806.4769 

\bibitem[Mohanty et al.(2005)]{2005MmSAI..76..303M} Mohanty, S., 
Jayawardhana, R., 
\& Basri, G.\ 2005, Memorie della Societa Astronomica Italiana, 76, 303 

\bibitem[Prosser et al.(1995)]{1995PASP..107..211P} Prosser, C.~F., et al.\ 
1995, Publications of the Astronomical Society of the Pacific, 107, 211 

\bibitem[Radick et al.(1987)]{1987ApJ...321..459R} Radick, R.~R., Thompson, 
D.~T., Lockwood, G.~W., Duncan, D.~K., 
\& Baggett, W.~E.\ 1987, Astrophysical Journal, 321, 459 

\bibitem[{{Rebull} {et~al.}(2006){Rebull}, {Stauffer}, {Megeath}, {Hora}, \&
  {Hartmann}}]{2006ApJ...646..297R}
{Rebull}, L.~M., {Stauffer}, J.~R., {Megeath}, S.~T., {Hora}, J.~L., \&
  {Hartmann}, L. 2006, Astrophysical Journal, 646, 297

\bibitem[{{Rebull} {et~al.}(2004){Rebull}, {Wolff}, \&
  {Strom}}]{2004AJ....127.1029R}
{Rebull}, L.~M., {Wolff}, S.~C., \& {Strom}, S.~E. 2004, Astronomical Journal, 127, 1029

\bibitem[Rebull et al.(2002)]{2002AJ....124..546R} Rebull, L.~M., Wolff, 
S.~C., Strom, S.~E., \& Makidon, R.~B.\ 2002, Astronomical Journal, 124, 546 

\bibitem[Rebull(2001)]{2001AJ....121.1676R} Rebull, L.~M.\ 2001, Astronomical Journal, 121, 
1676 

\bibitem[Reiners 
\& Basri(2008)]{2008ApJ...684.1390R} Reiners, A., \& Basri, G.\ 2008, Astrophysical Journal, 684, 1390 

\bibitem[Scholz 
\& Eisl{\"o}ffel(2007)]{2007MNRAS.381.1638S} Scholz, A., \& Eisl{\"o}ffel, J.\ 2007, Monthly Notices
of the Royal Astronomical Society, 381, 1638 

\bibitem[Scholz et al.(2007)]{2007ApJ...662.1254S} Scholz, A., Coffey, J., 
Brandeker, A., \& Jayawardhana, R.\ 2007, Astrophysical Journal, 662, 1254 

\bibitem[Scholz et 
al.(2005)]{2005A&A...438..675S} Scholz, A., Eisl{\"o}ffel, J., \& Froebrich, D.\ 2005, Astronomy \&
Astrophysics, 438, 675 

\bibitem[Scholz 
\& Eisl{\"o}ffel(2005)]{2005A&A...429.1007S} Scholz, A., \& Eisl{\"o}ffel, J.\ 2005, Astronomy \&
Astrophysics, 429, 1007 

\bibitem[Scholz 
\& Eisl{\"o}ffel(2004)]{2004A&A...421..259S} Scholz, A., \& Eisl{\"o}ffel, J.\ 2004, Astronomy \&
Astrophysics, 421, 259 

\bibitem[Scholz 
\& Eisl{\"o}ffel(2004)]{2004A&A...419..249S} Scholz, A., \& Eisl{\"o}ffel, J.\ 2004, Astronomy \&
Astrophysics, 419, 249 

\bibitem[Shu et al.(1994)]{1994ApJ...429..781S} Shu, F., Najita, J., 
Ostriker, E., Wilkin, F., Ruden, S., \& Lizano, S.\ 1994, Astrophysical Journal, 429, 781 

\bibitem[Sills et al.(2000)]{2000ApJ...534..335S} Sills, A., Pinsonneault, 
M.~H., \& Terndrup, D.~M.\ 2000, Astrophysical Journal, 534, 335 

\bibitem[Skumanich(1972)]{1972ApJ...171..565S} Skumanich, A.\ 1972, Astrophysical Journal, 
171, 565 

\bibitem[Stassun 
\& Terndrup(2003)]{2003PASP..115..505S} Stassun, K.~G., \& Terndrup, D.\ 2003, Publications
of the Astronomical Society of the Pacific, 115, 505 

\bibitem[Stassun et al.(2001)]{2001AJ....121.1003S} Stassun, K.~G., 
Mathieu, R.~D., Vrba, F.~J., Mazeh, T., \& Henden, A.\ 2001, Astronomical Journal, 121, 1003 

\bibitem[Stassun et al.(1999)]{1999AJ....117.2941S} Stassun, K.~G., 
Mathieu, R.~D., Mazeh, T., \& Vrba, F.~J.\ 1999, Astronomical Journal, 117, 2941 

\bibitem[Tinker et al.(2002)]{2002ApJ...564..877T} Tinker, J., 
Pinsonneault, M., \& Terndrup, D.\ 2002, Astrophysical Journal, 564, 877 

\bibitem[Zapatero Osorio et 
al.(2003)]{2003A&A...408..663Z} Zapatero Osorio, M.~R., Caballero, J.~A., B{\'e}jar, V.~J.~S., 
\& Rebolo, R.\ 2003, Astronomy \& Astrophysics, 408, 663 

\end{thebibliography}

\IfFileExists{\jobname.bbl}{}
 {\typeout{}
  \typeout{******************************************}
  \typeout{** Please run "bibtex \jobname" to optain}
  \typeout{** the bibliography and then re-run LaTeX}
  \typeout{** twice to fix the references!}
  \typeout{******************************************}
  \typeout{}
 }

\end{document}